\documentstyle[12pt]{article}  \topmargin -0.5in
\begin{document}
\begin{titlepage} \title{\begin{flushright} {\normalsize UB-ECM-PF
94/31 } \end{flushright} \vspace{2cm} {\Large \bf Amplification of the
scattering cross section due to non-trivial topology of the
spacetime}} \author{K. Kirsten\cite{hugo} and Yu. Kubyshin \cite{yk}\\
Department d'ECM, Facultat de F{\'{\char'20}}sica \\ Universitat de
Barcelona, Av. Diagonal 647, 08028 Barcelona \\ Spain }

\date{December 6, 1994}

\maketitle

\begin{abstract} In refs.~\cite{torus-scat,sphere-scat} it was
demonstrated that the total cross section of the scattering of two
light particles (zero modes of the Kaluza-Klein tower) in the
six-dimensional $\lambda \phi^{4}$ model differs significantly from
the cross section of the same process in the conventional $\lambda
\phi^{4}$ theory in four space-time dimensions even for the energies
below the threshold of the first heavy particle. Here the analytical
structure of the cross section in the same model with torus
compactification for arbitrary radii of the two-dimensional torus is
studied. Further amplification of the total cross section due to
interaction of the scalar field with constant background Abelian gauge
potential in the space of extra dimensions is shown. \end{abstract}

\end{titlepage}

\newtheorem{Def}{Definition} \newtheorem{lem}{Lemma}
\newtheorem{sat}{Satz}

\section{Introduction}

There are quite a few effects in quantum field theory due to
boundaries or non-trivial spacetime topology. Topological mass
generation and the Casimir effect are beautiful and simple
manifestations of them (see for example
\cite{top-mass}-\cite{Casimir}). Theories at non-zero temperature,
which are equivalent to those with the time dimension being curled up
to the circle with the radius (temperature)$ ^{-1}$, give another
physically interesting example of such effects (see
\cite{temperature}).

Scattering of particles in theories on the spacetime with non-trivial
topology was studied in ref. \cite{torus-scat}, \cite{sphere-scat} and
\cite{DIKT}. There a simple model of one scalar field $\phi$ with
quartic self-interaction on the spacetime of the type $M^{n} \times
K$, where $M^{n}$ is the $n$-dimensional Minkowski spacetime and $K$
is a two-dimensional compact space, was analyzed. As it is well known,
such a model can be re-written as an effective model on $M^{n}$ with
an infinite number of fields $\phi_{N}$, where $N$ is a multi-index
labelling the eigenfunctions of the Laplace operator $\Delta_{K}$ on
the manifold $K$. These fields have rising spectrum of masses given by
the formula of the type $M_{N}^{2} = m^{2} + N^{2}/L^{2}$, where $m$
is the mass of the original ($n+2$)-dimensional model and $L$ is a
characteristic size of the space $K$. Often the fields $\phi_{N}$ are
referred to as Kaluza-Klein modes and the infinite set of them is
called the Kaluza-Klein tower of modes. For the problems of physical
interest $Lm \ll 1$, so the mode with the mass $m$ is called the light
mode or zero mode, the rest are refered to as heavy modes. It is worth
to note that in the case of models with non-zero temperature $T$ in
the Euclidean time formalism the spacetime is described by $M^3 \times
S^{1}$. Then the scale $L$ is the radius of $S^1$ equal to $T^{-1}$
and the modes $\phi_{N}$ are labelled just by one integer ranging from
$-\infty$ to $+\infty$ and are usually called Matsubara modes.

The object of our study here is the total cross section of the
scattering process (2 light particles) $\rightarrow$ (2 light
particles). As one can easily see, in this model the heavy modes do
not contribute at the tree level. Thus, there is no difference between
the cross section $\sigma ^{(\infty)}$, calculated in the model
described above, and the cross section $\sigma ^{(0)}$, calculated in
the model on the spacetime $M^{n}$ (with the same mass $m$ and
corresponding coupling constants). It is the one loop order where
effects due to the non-trivial topology of the spacetime come into
play. It appears that because of contributions of virtual heavy
particles propagating along the loop the cross section has a behaviour
which differs significantly from that of $\sigma^{(0)}$ even for
energies of scattering particles noticeably below the threshold of the
first heavy particle. We should note that in models of another type
heavy modes can contribute already at the tree level, what makes the
effect stronger. Examples of such models appear in the superstring
theory with certain orbifold compactifications. Physical predictions
for some realistic processes and estimates on the size of $L$ were
obtained in \cite{antoniadis}.

The cases $n=2$, $K=T^{2}$ and $n=4$, $K=T^{2}$, where $T^{2} = S^{1}
\times S^{1}$ is the two-dimensional equilateral torus with the radius
$L$, with periodic boundary conditions were considered in
ref.~\cite{DIKT} and ref.~\cite{torus-scat} respectively. A detailed
analysis of the case of the spherical compactification, namely $n=4$,
$K=S^{2}$, with $L$ being the radius of the sphere, was carried out in
ref.~\cite{sphere-scat}. There it was also shown, that in the interval
of the centre of mass energies $\sqrt{s}$ of the particles below the
threshold of the first heavy mode, the contribution of the
Kaluza-Klein tower to the total cross section of the scattering is
essentially proportional to $\zeta(2| K) (sL^2)$, where $\zeta (2 |
K)$ is the zeta function  of the Laplace operator on the manifold $K$.
This is the way the topology of the spacetime enters into the
characteristic of the high energy process.

The calculations of the total cross section for $n=4$ are of physical
interest because in that case the model, in spite of its simplicity,
mimics some essential features of Kaluza-Klein type extensions of
physical models, for example of the Standard Model. One might think of
this type of extensions as low energy effective theories emerging, for
example, from superstrings.  Provided we are considering a realistic
model and the scale $L^{-1}$ is of the order of a few TeV (see below),
the effect can be (in principle) measured experimentally at future
colliders. This could yield evidence about the validity of the
Kaluza-Klein hypothesis within a given model.

Two important comments are in order here. The first one is, that the
scalar $\lambda \phi^{4}$ model in more than four dimensions, which is
considered in \cite{torus-scat} and \cite{sphere-scat} and which is
the object of the analysis in the present article, is
non-renormalizable. In the formalism used here this reveals in an
additional divergence of the sum over one-loop diagrams with heavy
modes propagating along the loop which contribute to the four-point
Green function or the scattering amplitude. Since the model is
regarded as a low energy effective theory coming from a more
fundamental theory, we do not consider the non-renormalizability to be
a principle obstacle. However, an additional prescription must be
imposed to treat this additional divergence. We choose it, as we
believe, in a physically acceptable way so as to make the difference
between the six dimensional model and the four dimensional one at the
scale of the four dimensional physics (i.e. at the energies of the
order of $m$ or smaller) as small as possible. If the prescription had
been imposed in some other way, then the difference between the models
would have been a low energy effect and could be observed immediately.
This would rule out the six-dimensional model from the very beginning.
It is worth mentioning that the six-dimensional models, considered
here, possess an additional property which makes their analysis more
interesting: the one-loop non-renormalizable contributions to the
total cross section of the two particle scattering process cancel out
on the mass shell when the $s$-, $t$- and $u$-channels are summed up.
Thus, we avoid the additional ambiguity due to non-renormalizability
in our calculations.

The second comment is related to the magnitude of the scale $L$,
characterizing new physics due to non-trivial topology of the original
model. In accordance with an analog of the decoupling theorem for this
class of theories \cite{decoupl} the effect naturally disappears if
the scale $L^{-1}$ is too large compared to the mass $m$. In many
approaches the scale of the compactification $L$ is assumed to be (or
appears to be) of the order of the inverse Planck mass $M_{Pl}^{-1}$
(see, for example, \cite{compact} and the reviews \cite{KK-review2}).
In this case additional dimensions could reveal themselves only as
peculiar gravitational effects or at an early stage of the evolution
of the Universe. On the other hand, there are some arguments in favour
of a larger compactification scale. One of them comes from
Kaluza-Klein cosmology and stems from the fact that the density of
heavy Kaluza-Klein particles cannot be too large, in order not to
exceed the critical density of the Universe. Estimates obtained in
ref. \cite{kolb} give the bound $L^{-1} < 10^{6}$ GeV. Other arguments
are related to results of papers \cite{kapl-88} and suggest that the
compactification scale should be of the order of the supersymmetry
breaking scale $M_{SUSY} \sim (1 \div 10)$ $TeV$. No natural mechanism
providing compactification of the space of extra dimensions with such
a scale is known so far. Having the above mentioned arguments in mind
we would like to study physical consequences in a multidimensional
model {\em assuming} that a compactification of this kind is indeed
possible.

In the present article we continue the analysis of the model described
above in the case of the spacetime $M^{4} \times T^{2}$. The aim is
twofold. First we carry out a more careful study of analytical
properties of the one-loop amplitude for positive and negative $p^{2}$
using the well developed machinery of zeta functions of the torus (see
for example \cite{eekk,kk}). The analysis will be extended to the case
of the non-equilateral torus and antiperiodic boundary conditions. The
second aim is to consider an extension of the model by coupling the
scalar field minimally to a constant Abelian gauge potential $A_{m}$
on the torus. Due to the non-trivial topology the constant components
$A_{m}$ cannot be removed by gauge transformations and are physical
parameters of the theory. We will show that the presence of such
classical gauge potential gives rise to an increase of the cross
section of the scattering of light particles for certain regions of
energies.

The paper is organized as follows. In Sect. 2 we describe the model,
choose the renormalization condition and discuss the general structure
of the 1-loop result for the four-point Green function and the total
cross section. In Sect. 3 detailed representations for the sum of
contributions of the Kaluza-Klein modes are derived. Behaviour of the
total cross section is analyzed in Sect. 4. Sect. 5 is devoted to the
analysis of the scattering cross section in the model with abelian
gauge potential. Conclusions and some discussion of the results are
presented in Sect. 6.

\section{Description of the model, mode expansion and renormalization}

We consider a one component scalar field on the $(4+2)$-dimensional
manifold $E=M^{4}\times T^{2}$, where $M^{4}$ is the Minkowski
space-time and $T^{2}$ is the two-dimensional torus of the radii
$L_{1}$ and $L_{2}$. In spite of its simplicity this model captures
some interesting features of both the classical and quantum properties
of multidimensional theories. The action is given by \begin{equation}
S= \int_{E} d^{4}x d^{2}y \left[\frac{1}{2}(\frac{\partial \phi
(x,y)}{\partial x^{\mu}})^2-\frac{1}{2}  \frac{\partial \phi (x,y)}
{\partial y^{i} } \frac{\partial \phi (x,y) }{\partial y^{j} } -
\frac{1}{2} m_{0}^{2} \phi^{2}(x,y) - \frac{\hat{\lambda}}{4!} \phi
^{4} (x,y) \right], \label{eq:action0} \end{equation} where $x^{\mu},
\mu = 0,1,2,3,$ are the coordinates on $M^{4}$, $y^{1}$ and $y^{2}$
are the coordinates on $T^{2}$, $0 < y^{1} < 2\pi L_{1}$, $0 < y^{2} <
2 \pi L_{2}$ and the field $\phi (x,y)$ satisfies periodic boundary
conditions on the torus. To re-interpret this model in
four-dimensional terms we make an expansion of the field $\phi (x,y)$,
\begin{equation}  \phi (x,y) = \sum_{N} \phi _{N} (x) Y_{N} (y),
\label{eq:expansion} \end{equation} where $N=(n_{1},n_{2})$, $-\infty
< n_{i} < \infty$ and $Y_{N}(y)$ are the eigenfunctions of the Laplace
operator on the internal space, \begin{equation}     Y_{(n_{1},n_{2})}
 =  \frac{1}{2 \pi \sqrt{L_{1} L_{2}}}     \exp \left[i \left(
\frac{n_{1}y^{1}}{L_{1}} +     \frac{n_{2}y^{2}}{L_{2}} \right)
\right]. \label{eq:laplace} \end{equation} Substituting this expansion
into the action and integrating over $y$, one obtains \begin{eqnarray}
   S & = & \int_{M^{4}} d^{4} x \left\{ \frac{1}{2} ( \frac{\partial
\phi_{0} (x)} {\partial x^{\mu}} )^{2} - \frac{1}{2} m_{0}^{2}
\phi_{0}^{2}(x) - \frac{\lambda_{1}}{4!} \phi _{0}^{4} (x) \right.
\nonumber  \\      &  & + \sum_{N>0}  \left[ \frac{\partial
\phi_{N}^{*} (x)}{\partial x^{\mu}} \frac{\partial \phi_{N}
(x)}{\partial x_{\mu}} - M_{N}^{2} \phi _{N}^{*} (x) \phi _{N} (x)
\right]  \nonumber  \\      &  & - \left. \frac{\lambda_{1}}{2} \phi
_{0}^{2} (x) \sum_{N > 0} \phi_{N}^{*}(x) \phi_{N}(x) \right\} -
S'_{int}, \label{eq:action1} \end{eqnarray} where the four-dimensional
coupling constant $\lambda_{1}$ is related to the multidimensional one
$\hat{\lambda}$ by $ \lambda_{1} = \hat{\lambda} / {\mbox volume}
(T^{2})$. In eq. (\ref{eq:action1}) the term $S_{int}'$ includes
vertices containing third and fourth powers of $\phi_{N}$ with $N^{2}
> 0$. We see that the model contains one real scalar field $\phi
\equiv \phi_{(0,0)}(x)$ describing a light particle of mass $m_{0}$,
and an infinite set (``tower") of massive complex fields $\phi_{N}(x)$
corresponding to heavy particles, or pyrgons, of masses given by
\begin{equation}   M_{N}^{2} = m_{0}^{2} + \frac{n_{1}^{2}}{L_{1}^2} +
  \frac{n_{2}^{2}}{L_{2}^2} .   \label{eq:mass} \end{equation}

Let us consider the 4-point Green function $\Gamma^{(\infty)}$ with
external legs corresponding to the light particles $\phi $. The index
$(\infty)$ indicates that the whole Kaluza-Klein tower of modes is
taken into account.

It is easy to see that the tree level contribution is the same as in
the dimensionally reduced theory whose action is given by the first
line in eq. (\ref{eq:action1}). At the one-loop level, owing to the
infinite sum of diagrams, the Green function to one-loop order is
quadratically divergent. This is certainly a reflection of the fact
that the original theory is actually six-dimensional and, therefore,
non-renormalizable. Thus, the divergencies cannot be removed by
renormalization of the coupling constant alone. We must also add a
counterterm $\lambda_{2B}\phi ^{2}(x,y) \Box _{(4+d)} \phi^{2}(x,y)$,
where $\Box _{(4+d)}$ is the D'Alembertian on $E$ and $\lambda_{2B}$
has mass dimension two. Of course, for the calculation of other Green
functions, or higher-order loop corrections, other types of
counterterms, which are not discussed here, are necessary. Hence, the
Lagrangian  we will use for our investigation is \begin{eqnarray}
{\cal L} & = & \frac{1}{2} ( \frac{\partial \phi _{0}(x)} {\partial x}
)^{2} - \frac{m_{0}^{2}}{2} \phi ^{2}_{0}(x)  \nonumber \\   & + &
\sum_{N > 0}  \left[ \frac{\partial \phi_{N}^{*} (x)}{\partial
x^{\mu}} \frac{\partial \phi_{N} (x)}{\partial x_{\mu}} - M_{N}^{2}
\phi _{N}^{*} (x) \phi _{N} (x) \right]        \nonumber \\   & - &
\frac{\lambda_{1B}}{4!} \phi _{0}^{4} (x) - \frac{\lambda_{1B}}{2}
\phi _{0}^{2} (x) \sum_{N > 0} \phi_{N}^{*}(x) \phi_{N}(x) -
\frac{\lambda_{2B}}{4!} \phi _{0}^{2}(x) \Box \phi _{0}^{2}(x),
\label{eq:action2} \end{eqnarray} where $\lambda _{1B}$ and
$\lambda_{2B}$ are bare coupling constants. To regularize the
four-dimensional integrals we will employ dimensional regularization,
which is performed, as usual, by making analytical continuation of the
integrals to $(4-2\epsilon)$ dimensions. $\kappa$ will be a mass scale
set up by the regularization procedure. The sum over the Kaluza-Klein
tower of modes will be regularized by means of the zeta function
technique \cite{hawk}.

The renormalization scheme is chosen in the same way as in ref.
\cite{torus-scat}, \cite{sphere-scat}. Let us refer the reader to
those articles for details and present here only the main formulas
which we will need for our calculations.

As subtraction point we choose the following point in the space of
invariant variables built up out of the external four-momenta  $p_{i}$
$(i=1,2,3,4)$  of the scattering particles: \begin{eqnarray} & &
p_{1}^{2}  =  p_{2}^{2} =   p_{3}^{2} =  p_{4}^{2} = m_{0}^{2},
\nonumber \\  & & p_{12}^{2} = s = \mu_{s}^{2}, \; \;      p_{13}^{2}
= t = \mu_{t}^{2}, \; \;      p_{14}^{2} = u = \mu_{u}^{2},
\label{eq:point} \end{eqnarray} where $p_{1j}^{2} = (p_{1} +
p_{j})^{2}$, $(j=2,3,4)$, and $s$, $t$ and $u$ are the Mandelstam
variables. Since the subtraction point is located on the mass shell,
it satisfies the standard relation $\mu_{s}^{2} + \mu_{t}^{2} +
\mu_{u}^{2} = 4 m_{0}^{2}$. The renormalization prescriptions for the
4-point Green function are as follows \begin{equation}  \left.
\Gamma^{(\infty)} (p_{1j}^{2}; m_{0}, L_{1},L_{2}, \lambda_{1B},
\lambda_{2B}, \epsilon) \right|_{s.p.} =   \left. \Gamma ^{(0)}
(p_{1j}^{2}; m_{0}, \lambda_{1B}', \epsilon) \right|_{s.p.}  =  g
\kappa ^{2\epsilon}, \label{eq:renorm1} \end{equation}
\begin{equation}  \left. \left[ \frac{\partial}{\partial p_{12}^{2}} +
\frac{\partial }{\partial p_{13}^{2}} +  \frac{\partial }{\partial
p_{14}^{2}}\right] \Gamma^{(\infty)} \right|_{s.p.} =  \left. \left[
\frac{\partial}{\partial p_{12}^{2}} + \frac{\partial }{\partial
p_{13}^{2}} +  \frac{\partial }{\partial p_{14}^{2}}\right]
\Gamma^{(0)} \right|_{s.p.} + \frac{\lambda_{2}}{4} \kappa ^{-2 +
2\epsilon}.     \label{eq:renorm2} \end{equation} Here $\Gamma ^{(0)}$
is the four-point Green function of the four-dimensional theory with
the zero mode field only (i.e., the dimensionally reduced theory),
$\lambda_{1B}'$ being its bare coupling constant. In the first line we
have written down the dependence of the Green functions on the
momentum arguments and parameters of the theory explicitly, and we
have taken into account that to one-loop order they depend on
$p_{12}^{2}$, $p_{13}^{2}$, and $p_{14}^{2}$ only. The label $s.p.$
means that the corresponding quantities are taken at the subtraction
point (\ref{eq:point}). $g$ and $\lambda_{2}$ are renormalized
coupling constants. The last one is included for the sake of
generality only, and we will see that our result does not depend on
it. More detailed discussion of the renormalization prescriptions
(\ref{eq:renorm1}), (\ref{eq:renorm2}) can be found in
\cite{sphere-scat}.

To one-loop order, the Green functions of the complete theory and of
the theory with only the zero mode are given by \begin{eqnarray}
\Gamma ^{(\infty)} (p_{1j}^{2}; m_{0}, L_{1},L_{2}, \lambda_{1B},
\lambda_{2B}, \epsilon) & = &\lambda_{1B} + \lambda_{2B}
\frac{p_{12}^{2}+p_{13}^{2}+p_{14}^{2}}{12} \nonumber \\  & + &
\lambda_{1B}^{2} [ K_{0}(p_{1j}^{2}; m_{0},\epsilon) +        \Delta K
(p_{1j}^{2}; m_{0},L_1,L_2,\epsilon) ], \label{eq:1loop} \\  \Gamma
^{(0)} (p_{1j}^{2}; m_{0}, \lambda_{1B}',\epsilon) & = &
\lambda_{1B}' + \lambda_{1B}'^{2} K_{0}(p_{1j}^{2}; m_{0}, \epsilon).
\nonumber \end{eqnarray} Here \begin{equation}
K_{0}(p_{1j}^{2};m_{0},\epsilon) \equiv K_{00}(p_{1j}^{2};m_{0}^{2},
\epsilon), \; \; \;  \Delta K (p_{1j}^{2};m_{0},L_{1},L_{2}
,\epsilon)= \sum_{N>0} K_{N}(p_{1j}^{2};M_{N}^{2},\epsilon)
\label{eq:Kexpansion} , \end{equation} and $K_{N}$ is the contribution
of the mode $\phi _{N}$ with the mass $M_{N}$ (see eq.
(\ref{eq:mass})) to the one-loop diagram of the scattering of two
light particles, \begin{equation} K_{N} (p_{1j}^{2};
M_{N}^{2},\epsilon) = \frac{-i}{32 \pi^{4}}
\frac{1}{M_{N}^{2\epsilon}} \left[ I(\frac{p_{12}^{2}}{M_{N}^{2}},
\epsilon) + I(\frac{p_{13}^{2}}{M_{N}^{2}}, \epsilon) +
I(\frac{p_{14}^{2}}{M_{N}^{2}}, \epsilon)\right].
\label{eq:Kdefinition} \end{equation} Here we assume that
$\lambda_{2B} \sim \lambda_{1B}^{2}$, so that the one loop diagrams
proportional to $\lambda_{1B} \lambda_{2B}$ or $\lambda_{2B}^{2}$ can
be neglected. It can be shown that this hypothesis is consistent (see
ref. \cite{decoupl}). As it stands, the function $\Delta K$is well
defined for $ \Re \epsilon > 1$, the sum being convergent for this
case. We will need, however, its value at $\epsilon = 0$ where it has
to be understood as the analytical continuation of
(\ref{eq:Kexpansion}). The same remark holds for all expressions of a
similar type appearing in the following. The function $I$ in the
formula above is the standard one-loop integral \begin{eqnarray}
I(\frac{p^{2}}{M^{2}},\epsilon )    & = & M^{2\epsilon} \int
d^{4-2\epsilon} q \          \frac{1}{(q^{2}+M^{2})((q-p)^{2}+M^{2})}
\nonumber \\    & = & i \pi^{2-\epsilon} \Gamma (\epsilon)
M^{2\epsilon} \int_{0}^{1} dx \ \frac{1}          {[M^{2} - p^{2}
x(1-x)]^{\epsilon}}. \label{eq:integral} \end{eqnarray} Let us also
introduce the sum of the one-loop integrals over all Kaluza-Klein
modes \begin{equation}   \Delta I
(p^{2}L_{1}^{2},m_{0}L_{1},w,\epsilon) = \sum_{N>0}
(\frac{1}{L_{1}^{2}M_{N}^{2}})^{\epsilon}
I(\frac{p^{2}}{M_{N}^{2}},\epsilon ), \label{eq:int-sum}
\end{equation} so that \begin{eqnarray}  \Delta K (p_{1j}^{2};
m_{0},L_{1},L_{2},\epsilon)    & = & \frac{-i}{32 \pi ^{4}}
L_{1}^{2 \epsilon}  \left[ \Delta I
(p_{12}^{2}L_{1}^{2},m_{0}L_{1},w,\epsilon)
\right.                \nonumber   \\      &+&  \left.      \Delta I
(p_{13}^{2}L_{1}^{2},m_{0}L_{1},w,\epsilon) +      \Delta I
(p_{14}^{2}L_{1}^{2},m_{0}L_{1},w,\epsilon) \right].
 \label{deltaK} \end{eqnarray} Here we denote $w = (L_{1}/L_{2})^{2}$.
Performing the renormalization, we obtain the following expression for
the renormalized four-point Green function \begin{eqnarray}   & \Gamma
^{(\infty)}_{R}&   \left(\frac{p_{1j}^{2}}{\mu_{j}^{2}};
\mu_{j}^{2}L_{1}^{2};m_{0}L_{1},w,g, \lambda_{2}\right)
                       \nonumber  \\   &=&  \lim _{\epsilon
\rightarrow 0} \kappa ^{-2 \epsilon}   \Gamma ^{(\infty)}
\left(p_{1j}^{2}; m_{0}, L_{1},L_{2},   \lambda_{1B} (g,\lambda_{2}),
 \lambda_{2B}(g, \lambda_{2} \right), \epsilon)  \nonumber \\   & = &
\lim _{\epsilon \rightarrow 0} \left\{ g + \lambda_{2}
\frac{p_{12}^{2}+p_{13}^{2}+p_{14}^{2}-\mu_{s}^{2}-\mu_{t}^{2}-
\mu_{u}^{2}}{12 \kappa^{2}}   \right.  \label{eq:gfren1}     \\   & +
& g^{2} \kappa ^{2 \epsilon} \left[ K_{0}(p_{1j}^{2};
m_{0},\epsilon) -   K_{0}(\mu_{j}^{2}; m_{0},\epsilon)  \right.
\nonumber \\   & + & \Delta K (p_{1j}^{2}; m_{0},L_{1},L_{2},\epsilon)
-   \Delta K (\mu_{j}^{2}; m_{0},L_{1},L_{2},\epsilon)
     \nonumber \\   & - & \left. \left. \left.   \frac{\sum_{j=2}^{4}
p_{1j}^{2}-\mu_{s}^{2}-   \mu_{t}^{2}-\mu_{u}^{2}}{3}  \sum_{j=2}^{4}
\frac{\partial }{\partial p_{1j}^{2}}    \Delta K (p_{1j}^{2};
m_{0},L_{1},L_{2},\epsilon)        \right|_{s.p.} \right] \right\},
\nonumber \end{eqnarray} where we denote $\mu_{2}^{2} = \mu_{s}^{2}$,
$\mu_{3}^{2} = \mu_{t}^{2}$ and $\mu_{4}^{2} = \mu_{u}^{2}$. The
r.h.s. of this expression is regular in $\epsilon$, and after
calculating the integrals and the sums over $N=(n_1,n_2)$ in $\Delta
K$ we take the limit $\epsilon \rightarrow 0$.

The above expression is rather general and valid for an arbitrary
subtraction point. For the four-point Green function of the complete
theory (i.e. with all the Kaluza-Klein modes) renormalized according
to the conditions (\ref{eq:renorm1}) and (\ref{eq:renorm2}) at the
subtraction point (\ref{eq:point}), and taken at a momentum point
which lies on the mass shell of the light particle \begin{eqnarray} &&
  \Gamma ^{(\infty)}_{R}
(\frac{s}{\mu_{s}^{2}},\frac{t}{\mu_{t}^{2}},\frac{u}{\mu_{u}^{2}};
\mu_{s}^{2}L_{1}^{2},\mu_{t}^{2}L_{1}^{2},
\mu_{u}^{2}L_{1}^{2};m_{0}L_{1},w, g)     \nonumber   \\   &   & \ \ \
= g + g^{2} \lim _{\epsilon \rightarrow 0}   \kappa ^{2 \epsilon} [
K_{0}(s,t,u;m_{0},\epsilon) -
K_{0}(\mu_{s}^{2},\mu_{t}^{2},\mu_{u}^{2}; m_{0},\epsilon) \nonumber
\\   &  &  \ \ \ +   \Delta K (s,t,u; m_{0},L_{1},L_{2},\epsilon) -
\Delta K (\mu_{s}^{2},\mu_{t}^{2},\mu_{u}^{2};
m_{0},L_{1},L_{2},\epsilon) ].   \label{eq:gfren2} \end{eqnarray} The
variables $s$, $t$ and $u$ are not independent, since they satisfy the
well known Mandelstam relation $s+t+u=4m_{0}^{2}$.

The formula above is rather remarkable. It turns out that on the mass
shell, due to cancellations between the $s$-, $t$- and $u$-channels,
the contribution proportional to $\lambda_{2}$ and the terms
containing derivatives of the one-loop integrals vanish. Thus, heavy
Kaluza-Klein modes contribute to the renormalized Green function on
the mass shell in exactly the same way as the light particle in the
dimensionally reduced theory does. Indeed, it can be easily checked
that the additional non-renormalized divergences, arising from the
infinite summation in $\Delta K$, cancel among themselves when the
three scattering channels are summed up together.

Next we calculate the total cross section $\sigma^{(\infty)}(s)$ of
the scattering process \ (2 {\em light particles}) \ $\longrightarrow
$ \ (2 {\em light particles}), \  in the case when the whole
Kaluza-Klein tower of heavy particles contribute. We compare it with
$\sigma ^{(0)} (s)$, which is the cross section in the dimensionally
reduced model, i.e. when only the light particle contributes. From the
discussion above it is clear that at low energies $\sigma
^{(\infty)}\approx \sigma^{(0)}$, so in what follows we take
$s>4m_0^2$.

The quantity which describes the deviation of the total cross section
from that in the four-dimensional theory due to the contributions of
the heavy particles is the following ratio: \begin{equation}  R
\left(\frac{sL_{1}^2}{4};\mu_{s}^{2}L_{1}^{2},
\mu_{u}^{2}L_{1}^{2},\mu_{t}^{2}L_{1}^{2};  m_{0}L_{1},w \right)
\equiv  16 \pi^{2} \frac{\sigma^{(\infty)}(s) -  \sigma^{(0)}(s)}{ g
\sigma^{(0)}(s)} .   \label{eq:delta-def} \end{equation} Using the
expression for the 4-point Green function (\ref{eq:gfren2}),
renormalized according to (\ref{eq:renorm1}) and (\ref{eq:renorm2}),
we calculate the corresponding total cross sections and obtain that,
to leading order (i.e. 1-loop order) in the coupling constant $g$, the
function (\ref{eq:delta-def}) is equal to \begin{eqnarray}  &&  R
\left(\frac{sL_{1}^2}{4};\mu_{s}^{2}L_{1}^{2},
\mu_{u}^{2}L_{1}^{2},\mu_{t}^{2}L_{1}^{2};  m_{0}L_{1},w \right)
                   \nonumber    \\  \ \ \ \ &  & = - \frac{i}{\pi^{2}}
 \lim_{\epsilon \rightarrow 0} (L_{1} \kappa)^{2 \epsilon}  \left\{
\Re \Delta I \left(sL_{1}^{2},m_{0}L_{1},w,\epsilon \right)   - \Delta
I \left(\mu_{s}^{2}L_{1}^{2},m_{0}L_{1},w,\epsilon \right)
\right.\nonumber \\  \ \ \ \ &  & + \frac{2}{s-4m_{0}^{2}} \int_{-(s-4
    m_{0}^{2})}^{0} du     \Delta I
\left(uL_{1}^{2},m_{0}L_{1},w,\epsilon \right)    - \Delta I
\left(\mu_{u}^{2}L_{1}^{2},m_{0}L_{1},w,\epsilon \right)
\nonumber \\  \ \ \ \ &&  - \left.     \Delta I
\left(\mu_{t}^{2}L_{1}^{2},m_{0}L_{1},w,\epsilon \right)    \right\}.
\label{eq:delta-exp} \end{eqnarray} Here we assume that
\begin{equation} \mu_{s}^{2},\mu_{t}^{2},\mu_{u}^{2} < 4 m_{0}^{2}.
\label{mu-condition} \end{equation}

\section{Calculation of the 1-loop contribution}

In this section we will analyze the 1-loop contribution of the heavy
Kaluza-Klein modes. The starting point is the expression
(\ref{eq:int-sum}). More detailed, the relevant expression is
\begin{equation} \zeta (\epsilon;
b^2,c^2,w)=\int\limits_0^1dx\,\,{\sum_{n_{1},n_{2}=
-\infty}^{\infty}}\!\!\!\!\!' \left[n_{1}^2+w n_{2}^2+c^{2} -
b^{2}x(1-x) \right]^{-\epsilon},                            \label{k2}
\end{equation} which is related to $\Delta I$ by \begin{equation}
\Delta I (p^{2}L_{1}^{2},m_{0}L_{1},w,\epsilon) =  i \pi^{2-\epsilon}
\Gamma (\epsilon)  \zeta (\epsilon; p^{2}L_{1}^{2},m_{0}^2 L_{1}^2,w),
                   \label{deltaI-zeta} \end{equation} and
$w=(L_1/L_2)^2$ was already introduced before. A detailed knowledge of
the behaviour of $\zeta (\epsilon ;b^2,c^2,w)$ as a function of $b^2$
around $\epsilon = 0$ is necessary. Specifically for the calculation
of the cross section and the function (\ref{eq:delta-exp}) we need an
analytical expression of eq.~(\ref{k2}) for positive and negative
$b^2$. For this two cases different techniques are needed and we will
present the two calculations one after the other.

Let us fix $b^2>0$ and start with $\zeta (\epsilon ;-b^2,c^2,w)$. Then
the effective mass term in eq.~(\ref{k2}), which is now $c^2 + b^2
x(1-x)$ is always greater than $0$ (we choose $c^2>0)$. For that case
it is very useful to perform re-summations, employing for
$t\in\mbox{${\rm I\!R }$} _+$, $z\in \mbox{${\rm I\!\!\!C }$}$ the
identity \cite{hille} \begin{equation} \sum_{n=-\infty}^{\infty}
\exp\{-tn^2+2\pi inz\}=\left(\frac{\pi} t\right)^{\frac 1
2}\sum_{n=-\infty}^{\infty}\exp\left\{-\frac{\pi^2} t
(n-z)^2\right\},\label{k3} \end{equation} which is due to Jacobi's
relation between theta functions. Using this and an integral
representation of the McDonald functions \cite{grad} (for details see,
for example, \cite{eekk,kk}), one finds the representation
\begin{eqnarray} \lefteqn{  \zeta(\epsilon ;-b^2,c^2,w)
=\frac{\pi}{\sqrt{w}}\frac 1 {\epsilon -1}\int\limits_0^1dx\,\, \left[
c^2+b^2x(1-x)\right]^{1-\epsilon}-\int\limits_0^1dx\,\,\left[
c^2+b^2x(1-x)\right]^{-\epsilon}  } \nonumber\\ &
&+\frac{\pi^{\epsilon}}{\sqrt{w}}\frac 2 {\Gamma (\epsilon )}
{\sum_{l,n=-\infty}^{\infty}\!\!\!^{\prime}}\int\limits_0^1dx\,\, \left[
c^2+b^2x(1-x)\right]^{\frac{1-\epsilon} 2}
\left[l^2+wn^2\right]^{\frac{\epsilon-1} 2}\times\nonumber\\ &
&\qquad\qquad K_{1-\epsilon}\left(2\pi \left[
c^2+b^2x(1-x)\right]^{\frac 1 2}[l^2+wn^2]^{\frac 1 2}\right)
\label{k4}   \\ & & = - \frac{\pi}{\sqrt{w}} \left( c^2 +
\frac{b^{2}}{6} \right) - 1    + {\cal O} (\epsilon).
\label{k4prime} \end{eqnarray} The advantage of eq.~(\ref{k4}) is,
that it consists of different contributions with a completely
different behaviour for large values of $b^2$. So it may be shown,
that the contributions of the McDonald functions, even though there is
a double sum, is negligable against the first terms due to the
exponentially fall off of the McDonald functions for large argument.

To obtain the cross section and the ratio (\ref{eq:delta-exp}) we need
the integral of the function (\ref{k4}): \begin{equation} S(s;\epsilon
)=\frac{1}{s} \int\limits_{0}^{s}du\,\,           \zeta (\epsilon
;-uL_1^2,a^2,w).            \label{k5} \end{equation} Doing first the
$u$-integration and continuing like in the calculation of
eq.~(\ref{k4}), a similar representation for $S(s;\epsilon )$ can be
found. We need only the first two terms of its Taylor expansion in
$\epsilon$ at $\epsilon = 0$. They read \begin{eqnarray}  S(s;
\epsilon) & = & S(s;0) + \epsilon S'(s;\epsilon = 0) +
{\cal O} (\epsilon ^2),   \nonumber  \\  S(s; 0) & = & -\frac{\pi}{2
\sqrt{w}} \left( 2 c^2 + \frac{s L_{1}^2}                 {6} \right)
-1       \label{k6prime} \end{eqnarray} and \begin{eqnarray} \lefteqn{
S'(s; 0)=\frac{\pi}{2s\sqrt{w}L_1^2}\int\limits_0^1\frac{dx}{x(1-x)}\left\{[
c^2+sL_1^2x(1-x) ]^2 \ln [ c^2+sL_1^2x(1-x) ]\right.   }\nonumber\\ &
&\left.-c^4\ln c^2+\frac 3 2 c^4-\frac 3 2 [ c^2+sL_1^2x(1-x)
]^2\right\}
+\frac{1}{sL_1^2}\int\limits_0^1\frac{dx}{x(1-x)}\left\{-sL_1^2x(1-x)
\right.     \label{k6} \\ & & \left.    +[ c^2+sL_1^2x(1-x) ]\ln [
c^2+sL_1^2x(1-x) ]-c^2\ln c^2\right\}         \nonumber\\ & &+\frac
{2} {\pi s
\sqrt{w}L_1^2}\int\limits_0^1\frac{dx}{x(1-x)}{\sum_{l,n=-\infty}^{\infty}\!\!\!
   ^{\prime}}
[l^2+wn^2] ^{-1}\left\{c^2K_2\left(2\pi c[l^2+wn^2]^{\frac 1 2}\right)
               \right.             \nonumber\\ & & \left. -[
c^2+sL_1^2x(1-x)]K_2\left(2\pi [  c^2+sL_1^2x(1-x) ]^{\frac 1 2}
[l^2+wn^2]^{\frac 1 2}\right)\right\}. \nonumber \end{eqnarray}

It may be seen, that for finite values of $s$ this expression is well
defined. Apart from the contributions including the McDonald
functions, all integrations are elementary \cite{grad}. However, the
result is even longer than the one presented and we will not do so
explicitly. Instead, let us only mention, that the leading behaviour
for $s\to \infty$ is \begin{equation} S'(s\to \infty ; \epsilon =0)
=\frac{\pi}{6 \sqrt{w}}sL_1^2\left\{-\frac{11}{6} +\ln
(sL_1^2)\right\}+ {\cal O} (sL_1^2\ln (sL_1^2))          \label{k7}
\end{equation}

Up to now, the presented results were derived for $-b^2$, that is for
$p^2 < 0$, especially useful for large $b^2$. When one tries to use
the presented techniques for $\zeta (\epsilon; b^2,c^2,w)$ with
$b^2>0$, one directly encounters infinities. That there must be
problems is seen in eq.~(\ref{k4}), because the argument of the
McDonald functions then lies on its cut.

Thus we have to proceed in a different way which is to use the
binomial expansion method. This method will yield suitable results for
small values of $b^2$, valid independent of its sign. The result is
given as a power series in $b^2$, the radius of convergence is
determined by the first mass in the Kaluza-Klein tower. Using this
method we get \begin{eqnarray} \zeta (\epsilon; b^2,c^2,w)&=&
\int\limits_0^1dx\,\,
{\sum_{n_{1},n_{2}=-\infty}^{\infty}\!\!\!\!\!\!^{\prime}}
[n_{1}^2+wn_{2}^2+c^2]^{-\epsilon}\left\{1-\frac {b^2x(1-x)}
{n_{1}^2+wn_{2}^2+c^2}\right\}^{-\epsilon}\nonumber\\
&=&\sum_{k=0}^{\infty}\frac{\Gamma(\epsilon +k)}{\Gamma(\epsilon
)}\frac{k!}{(2k+1)!}Z_2^{c^2}(\epsilon +k; 1,w)b^{2k},  \label{k8}
\end{eqnarray} where we introduced the Epstein-type zeta-function
\begin{equation} Z_2^{c^2}(\nu
;u_1,u_2)={\sum_{n_{1},n_{2}=-\infty}^{\infty}\!\!\!\!\!\!^{\prime}}
[u_1 n_{1}^2+u_2 n_{2}^2+c^2]^{-\nu}.       \label{k9} \end{equation}
It can be shown that this function has the following properties:
\begin{eqnarray}  Z_{2}^{c^2}(\epsilon;1,w) & = &               -
\left( \frac{\pi c^2}{\sqrt{w}} + 1 \right) +       \epsilon
{Z'}^{c^2}_2(0;1,w)+{\cal O}(\epsilon ^2),   \nonumber \\
Z_{2}^{c^2}(1+\epsilon;1,w) & = &                \frac{\pi}{\sqrt{w}}
\frac{1}{\epsilon} + PP\,\,Z_2^{c^2}(1;1,w)+{\cal O}(\epsilon ).
   \label{k10} \end{eqnarray} Using these formulas we obtain that the
expression (\ref{k8}) reads \begin{eqnarray} \zeta (\epsilon;
b^2,c^2,w)&=&          \zeta(0;b^2,c^2,w) + \epsilon
\zeta'(0;b^2,c^2,w)   + {\cal O} (\epsilon^2)  \nonumber  \\
&=& - \frac{\pi}{\sqrt{w}} \left(c^2 - \frac{b^2}{6} \right)   - 1
             \nonumber \\    &+& \epsilon \left[{Z'}_2^{c^2}(0;1,w)
+\frac{b^2} 6 PP\,\,Z_2^{c^2} (1;1,w)\right]\nonumber\\   & + &
\epsilon \sum_{k=2}^{\infty}\frac{\Gamma(k)k!}   {(2k+1)!}Z_2^{c^2}(k;
1,w)b^{2k}.   \label{k10prime} \end{eqnarray} Expressions for
${Z'}_2^{c^2}$ and the finite part $PP\,\,Z_2^{c^2}$ are rather
lengthy \cite{eekk}. We need not to present them here explicitly,
because for the calculation of our main objective $R$,
eq.~(\ref{eq:delta-exp}), contributions $\sim p^2$ and constant in
$p^2$ cancel out as it was explained in Sect.~2.

We observe that in the limit $\epsilon \rightarrow 0$ (\ref{k10prime})
coincides with (\ref{k4prime}). This ensures that after substituting
eqs. (\ref{k4prime}), (\ref{k6prime}), (\ref{k6}) and (\ref{k10prime})
into (\ref{eq:delta-exp}) all terms singular in $\epsilon$ cancel so
that $R$ is regular at $\epsilon = 0$. For this cancellation it is
important that we choose the subtraction point to be on the mass
shell, i.e. $\mu_{s}^2+\mu_{u}^2+\mu_{t}^2 =m_{0}^2$, as it was
discussed in Sect. 2. The finite part of (\ref{eq:delta-exp}) is
calculated using eq. (\ref{k6}) for the integral term and eq.
(\ref{k10prime}) for the rest of the terms.

As mentioned, the representation (\ref{k8}) is valid up to the first
threshold. This is seen as follows. In eq.~(\ref{k10prime}) we need
the behaviour for $k\to \infty$. Without loss of generality let us
assume $w=(L_1/L_2)^2 \geq 1$. In the considered limit the smallest
summation indices are only important and we find \begin{equation}
Z_2^{c^2}(k; 1,w) \sim  m_{1}(w) [1+c^2]^{-k},   \label{k10bis}
\end{equation} where $m_{1}(w)$ is the multiplicity of the first heavy
mode in accordance with eq.~(\ref{eq:mass}). The condition of
convergence then reads \begin{equation} \frac{b^2}{4} < 1+c^2.
  \label{k11} \end{equation} This means that for the first term in the
r.h.s. of (\ref{eq:delta-exp}) the representation (\ref{k8}) is valid
up to the threshold of the first heavy particle, i.e. up to $s < 4
M_{(1,0)}^2 = 4 (m_{0}^2 + 1/L_{1}^2)$. Here we assume that the
subtraction points are chosen to satisfy (\ref{mu-condition}), so that
the terms in (\ref{eq:delta-exp}) evaluated at the subtraction points
converge.

In this calculation formally also any manifold of the kind $M^4\times
K$, with $K$ an arbitrary compact two-dimensional manifold may be
dealt with. The only difference is that in the results (\ref{k10}) the
Epstein-type zeta-function has to be replaced by the corresponding one
of $K$.

Of course the question arises, how one may obtain a similar
representation for the cross section extended beyond the first
threshold. One needs to find an analytical continuation of
eq.~(\ref{k8}), or more detailed for $\zeta'(0;b^2,c^2,w)$, for values
$(|b^2|/4)\geq 1+c^2$. The behaviour of the sum near $(|b^2|/4) \sim
(1+c^2)$ in eq. (\ref{k10prime}) is determined by the behaviour of the
function \cite{grad} \[ \sqrt{1-x^2}\arctan\frac x
{\sqrt{1-x^2}}=x-\frac 1 4 \sum_{k=1}^{\infty}
\frac{\Gamma(k)k!}{(2k+1)!}(2x)^{2k+1} \] near $|x|=1$ (cf.
(\ref{k10prime}), (\ref{k10bis})). Subtracting it from and adding it
to eq.~(\ref{k10prime}) with $x^2=b^2/[4(1+c^2)]$, we get
\begin{eqnarray} \lefteqn{ \zeta'(0;b^2,c^2,w) =
\sum_{k=2}^{\infty}(-1)^k\frac{(k-1)!k!}{(2k+1)!}Z_2^{c^2}(k;1,w)b^{2k}
}\nonumber\\ &=&2 m_{1}(w)\left\{1-\frac 1 3 \frac{x^2}3
-\frac{\sqrt{1-x^2}} x \arctan \frac x
{\sqrt{1-x^2}}\right\}\nonumber\\ &
&+\sum_{k=2}^{\infty}\frac{(k-1)!k!}{(2k+1)!} \left[Z_2^{c^2}(k;1,w)
-\frac {m_{1}(w)}{(1+c^2)^k} \right]b^{2k}.    \label{artur}
\end{eqnarray} The advantage of this representation is apparent. The
sum in eq.~(\ref{artur}) is convergent up to the second threshold. The
remaining terms contain explicitly the analytical behaviour of
$\zeta'(0;b^2,c^2,w)$ at the first threshold. This is seen very well
by means of the formula $\ln (ix+\sqrt{1-x^2})=i\arctan
x/\sqrt{1-x^2}$, which may be used in eq.~(\ref{artur}) as well and
provides the analytical continuation of $\zeta'$ in $x^2$ up to the
second threshold. It is clear how to continue the procedure in order
to obtain a representation in principle valid up to any given
threshold.

Using representation (\ref{artur}) of $\zeta'$, the function
$S'(s;0)$, see eq.~(\ref{k5}), which is also important for the
calculation of the ratio $R$, see eq.~(\ref{eq:delta-exp}), may be
written in the form \begin{eqnarray} S'(s;0)&=&2 m_{1}(w)\left[1+\frac
1 6 \frac{sL_1^2}{4(1+c^2)}\right]-
m_{1}(w)F\left(\frac{sL_1^2}{4(1+c^2)}\right)\nonumber\\ &
&+\sum_{k=2}^{\infty}\frac{(k-1)!k!}{(2k+1)!} \left[Z_2^{c^2}(k;1,w)
-\frac {m{1}(w)}{(1+c^2)^k} \right](sL_1^2)^{2k},\label{neuartur}
\end{eqnarray} with \begin{equation} F(z) =-1 +2\sqrt{\frac{1+z} z}
\ln(\sqrt{1+2}+\sqrt{z})+\frac 1 z (\ln
(\sqrt{1+z}+\sqrt{z}))^2.\nonumber \end{equation} The presented
formulas (\ref{artur}) and (\ref{neuartur}) will appear to be quite
effective for the calculation of the ratio $R$ in the next section.

\section{Numerical results for the scattering cross section}

Using representations (\ref{k6prime}) and (\ref{k10prime}) we write
the ratio $R$ as \begin{eqnarray} \lefteqn{  R
\left(\frac{sL_{1}^2}{4};\mu_{s}^{2}L_{1}^{2},
\mu_{u}^{2}L_{1}^{2},\mu_{t}^{2}L_{1}^{2};  m_{0}L_{1},w \right)  =
}\nonumber\\ & &+\Re \left\{ \zeta'(0;sL_{1}^2,m_{0}^2L_{1}^2,w) -
\zeta'(0;\mu_{s}^2 L_{1}^2,m_{0}^2L_{1}^2,w)    \right.\nonumber \\
& & + \left. 2 S'(s-4m_{0}^2;0) -   \zeta'(0;\mu_{u}^2
L_{1}^2,m_{0}^2L_{1}^2,w)    - \zeta'(0;\mu_{t}^2
L_{1}^2,m_{0}^2L_{1}^2,w)  \right\}, \label{n1} \end{eqnarray} where
$S'$ is given by eq. (\ref{k6}) or (\ref{neuartur}) and $\zeta'$ is
given by (\ref{k10prime}) or (\ref{artur}) depending on the range of
the energy of the colliding particles.

For the numerical computation we take the zero mode particle to be
much lighter than the first heavy mode and choose the subtraction
points to be at the low energy interval. We take \begin{equation}
m_{0}^2 L_{1}^2 = 10^{-4}, \; \; \; \; \mu_{s}^2 / m_{0}^2 = 10^{-2},
\; \; \; \; \mu_{u}^2 = \mu_{t}^2 = (4m_{0}^2 - \mu_{s}^2)/4.
\label{n2} \end{equation} By making such a choice of parameters we
were motivated by the arguments in favour of the possibility of having
the compactification scale to be of the order of the supersymmetry
breaking scale, $L_{1}^{-1} \sim M_{SUSY}$ (see discussion in the
Introduction). Then the values (\ref{n2}) could mimic a situation
with, for example, $m_{0} = 100$ GeV, $L_{1}^{-1} = 10$ TeV.

Now, we assume that $w=(L_{1}/L_{2})^2 \geq 1$ and compute $R$ as a
function of $z=s/(4M_{(1,0)}^2)$, where due to our choice of $w$ we
have $M_{(1,0)}^2 = 1/L_{1}^2 + m_{0}^2$ for the square of the mass of
the first heavy mode. An approximate expression for the ratio with the
parameter values $m_0L_1^2,\mu_s^2L_1^2,\mu
_u^2L_1^2,\mu_t^2L_1^2\approx 0$ is easily obtained to be
\begin{equation} R(z;w) = \frac{4}{9} Z_2^{c^2}(2;1,w) z^2 +
\frac{8}{105} Z_2^{c^2}(3;1,w) z^3 + {\cal O} (z^4) \label{ratio1}
\end{equation} for $0 \leq z < 1$. Here we have supressed a part of
the arguments of the function (\ref{eq:delta-def}): $R(z;w) \equiv
R(z;0,0,0,0,w)$. To have an expression for $R$ valid at the first
threshold and above up to the second threshold the formula
(\ref{artur}) should be used. Then we get \begin{eqnarray} R(z;w) &=&
m_{1}(w)\left[6-2\sqrt{\frac{1-z} 2}\arctan \sqrt{\frac z {1-z}}
-2F(z)\right]\label{ratio}\\ &+&\frac 4 9
\left(Z_2^{c^2}(2;1,w)-m_{1}(w)\right)z^2+\frac 8 {105} \left(
Z_2^{c^2}(3;1,w)-m_{1}(w)\right) z^3 + {\cal O} (z^4).\nonumber
\end{eqnarray}

Plots of $R$ for various values of $w$ are presented in Fig.~1. Let us
first consider the interval $0 < z \leq 1$, i.e.~below the first
threshold. Even in this range the deviation of the cross section of
the theory on $M^4\times T^2$ from that of the four dimensional one,
characterized by $R$, is quite noticeable. Thus, for example, for
$w=1$ we find $R\approx 0.76$ for $z=0.5$ and $R\approx 0.17$ for
$z=0.25$. We would like to mention that the case $w=1$ was first
studied in \cite{torus-scat}. The closer is the space of extra
dimensions $T^2$ to the equilateral torus with $w=1$ the stronger is
the deviation of the cross section from the four dimensional one. This
might be the most relevant case, because due to the high symmetry of
this compactification the vacuum energy of the spacetime probably
takes a minimum value (for example this result has been found for a
scalar field living on a torus \cite{ambjorn}).

{}From the first line in eq. (\ref{ratio}) we see that for $0 < z \leq
1$ the behaviour of $R$ is basically determined by the multiplicity
$m_{1}(w)$ times some universal function of $z$. Since $m_{1}(1)=4$
and $m_{1}(w>1)=2$, this explains that the quotient $R(z;1)/R(z;w)
\approx 2$, as it can be seen from Fig. 1. The second line gives
corrections depending on $w$. With increasing $w$ these corrections
are getting smaller and for values of $w\geq 10$ they are already
negligable. On the contrary, for higher values of $z$, $z\geq 1$, they
are getting more important and as is seen in Fig.~1 the observations
true for $0\leq z<1$ are not true any more. This indicates that the
contribution of the heavy modes to the total cross section of the (2
light particles) $\to $ (2 light particles) scattering process grows
with the centre of mass energy $\sqrt{s}$ keeping the radii of the
compactification fixed.

The Fig.~2 shows the behaviour of $R$ as a function of the scale $L_1$
with $\sqrt{s}$ being fixed. As implemented by our renormalization
condition, eq.~(\ref{eq:renorm1}), (\ref{eq:renorm2}), for small $L_1$
we have $R\to 0$. Here we see once more, as already mentioned, that
for bigger centre of mass energy the influence of the heavy modes is
increasing. The setting described in Fig.~2 would be more appropriate
for making possible predictions in high energy experiments at modern
colliders (of course in more realistic models). So, in case $\sigma
(s)$ is measured experimentally and a value of $R =(\sigma (s) -\sigma
^{(0)} (s) )/\sigma ^{(0)} (s) \neq 0$ is found, Fig.~2 could be used
to obtain bounds on the size $L_1$ of extra dimensions.

We see that there is a noticeable deviation of the cross section
$\sigma^{(\infty)}$ of the complete theory from the cross section
$\sigma^{(0)}$ of the four-dimensional model, characterized by the
function $R$, due to the presence of the heavy Kaluza-Klein modes. The
maximal "amplitude" of this deviation below or at the threshold of the
first heavy particle is basically determined by the multiplicity of
this mode. Another physical situation illustrating this property is
considered in the next section.

\section{Am\-pli\-fi\-ca\-tion of the cross sec\-tion by con\-stant
abe\-li\-an gauge field}

In generalization of the model described by the action
(\ref{eq:action0}), let us now consider the abelian scalar gauge
theory on $M^{4} \times T^{2}$. We will be interested in the case when
the only non-zero components of the abelian gauge potential are those
of the extra dimensions and, moreover, here we will consider them as
classical external fields. With these assumptions the action of the
theory we are going to study is \begin{eqnarray} S&=& \int_{E} d^{4}x
d^{2}y \left[\frac{1}{2}(\frac{\partial\phi (x,y)} {\partial
x^{\mu}})^2+\frac{1}{2}\left[ \left( \frac{\partial}{\partial y
^i}-A_i\right) (x,y)\right]\left[  \left( \frac{\partial}{\partial y
^i}-A_i\right) \phi (x,y)\right]\right.\nonumber\\ &
&\qquad\qquad\left. - \frac{1}{2} m_{0}^{2} \phi ^{2}(x,y) -
\frac{\hat{\lambda}}{4!} \phi ^{4} (x,y) \right], \label{k13}
\end{eqnarray} with periodic boundary conditions for $\phi(x,y)$ in
the toroidal directions, $\phi(x^{\mu},y^{i}+2\pi L_{i}) =
\phi(x^{\mu},y^i)$, as before. For some previous studies of gauge
theories on the torus see \cite{gauge-torus}, \cite{actor}.

The model has some properties in common with theories with the abelian
gauge field at finite temperature $T$ (see, for example, \cite{actor},
\cite{smilga} for rewiews and references therein). Thus, in such
theories
the gauge potential component $A_{0}$ along the compact Euclidean time
direction becomes an angular variable in the effective action, i.e.
locally $- \pi \leq A_{0}/T \leq \pi$. In our model, as we will see
shortly, the function $R$ is periodic in the variables $A_{i}L_{i}$,
$(i=1,2)$. Also, in our model, similar to the theories at finite $T$
\cite{angular}, due to the fact that $T^2$ is a multiply-connected
space, the $A_{i}$'s
{\em cannot} be gauged away and are physical parameters of the model.
This is similar to the appearence of non-integrable phases of Wilson
line integrals as physical degrees of freedom in non-abelian gauge
theories on multiply-connected spaces \cite{hosotani}, \cite{weiss}.

Our intention here is to study the change of the function $R$ due to
the change of the spectrum of the heavy masses, namely the values of
$M_{(n_{1},n_{2})}$ and their multiplicities, produced by the gauge
potential. The ideal configuration to use for this is $A_{\mu}=0$
(this choice is already done in eq. (\ref{k13})) and $A_{i}=$const.
This is the approximation of the same type which is usually used for
studies in the theories at non-zero $T$. A more general configuration
for the abelian gauge potential on $M^{4} \times T^{2}$ would lead to
the model with the abelian gauge field, massive vector fields and
additional massive scalar fields on $M^{4}$, which is beyond the scope
of our investigation.

Substituting the Fourier expansion (\ref{eq:laplace}) into the action
(\ref{k13}) and integrating over the toroidal components we obtain the
model with the action (\ref{eq:action1}) but now the masses of the
fields depend on the gauge field components and are given by
\begin{equation}   M_{N}^2 (a) = m_{0}^2 +
\frac{(n_{1}-a_{1})^2}{L_{1}^2}   + \frac{(n_{2}-a_{2})^2}{L_{2}^2},
\label{k14} \end{equation} where $a_{i}=A_{i} L_{i}$, $i=1,2$. Notice
that now the mass of the zero mode, given by \begin{equation}
M_{(0,0)}^2 (a) = m_{0}^2 + \frac{a_{1}^2}{L_{1}^2}   +
\frac{a_{2}^2}{L_{2}^2},  \label{k15} \end{equation} also receives a
contribution from the gauge field components. This makes the
separation into the light masses and the heavy ones rather
problematic. We will return to this issue shortly.

We calculate then the cross section of the scattering of two {\em
zero} modes. Imposing the same renormalization conditions as
eqs.~(\ref{eq:renorm1}), (\ref{eq:renorm2}) and repeating all steps of
the calculations of Sects.~2 and 3 we obtain the following analogous
expression for the total cross section: \begin{eqnarray}  \sigma
^{(\infty)} (s,a) & = & g^2 + \frac{i g^3}{16 \pi^4}  \lim _{\epsilon
\rightarrow 0} (L\kappa)^{2\epsilon}\times\nonumber\\ & & \left\{
\frac{32 \pi^4 i}{L^{2 \epsilon}} \left[
K_{0}(s,t,u;M_{(0,0)}(a),\epsilon)  -
K_{0}(\mu_{s}^2,\mu_{t}^2,\mu_{u}^2;M_{(0,0)}(a),\epsilon)  \right]
\right.                   \nonumber \\  & & +\Re \Delta I \left(
sL^2,m_{0}L,a_{i};\epsilon \right)  - \Delta I \left( \mu_{s}^2
L^2,m_{0}L,a_{i};\epsilon \right)          \label{k16}  \\  & & +
\frac{2}{s-4m_{0}^2} \int_{-(s-4m_{0}^2)}^{0}  du \Delta I \left(
uL^2,m_{0}L,a_{i};\epsilon \right)\nonumber\\ & &  \left.- \Delta I
\left( \mu_{u}^2 L^2,m_{0}L,a_{i};\epsilon \right) - \Delta I \left(
\mu_{t}^2 L^2,m_{0}L,a_{i};\epsilon \right)   \right\},    \nonumber
\end{eqnarray} where $K_{0}$ is defined by eq.~(\ref{eq:Kexpansion})
and $\Delta I$ is given by eq. (\ref{eq:int-sum}) with $M_{N}^2$ being
replaced by (\ref{k14}). Here we restrict ourselves to the case of the
equilateral torus with $L_{1}=L_{2}=L$, so we supressed the dependence
on $w$ and instead indicated explicitly the dependence on the gauge
field parameters $a_{i}$. The analog of the function (\ref{k2}) is \[
\zeta ^A(\epsilon; b^2,c^2)=\int\limits_0^1dx\,\,{\sum_{n_{1},n_{2}
=-\infty}^{\infty}}\!\!\!\!\!' \left[(n_{1}-a_1)^2+(n_{2}-a_2)^2+
c^2+b^2x(1-x)\right]^{-\epsilon}. \] The total cross section
(\ref{k16}) includes summation over {\em all} modes and is periodic in
$a_{i}$: \begin{eqnarray} \sigma
^{(\infty)}(s,a_{1}+k_{1},a_{2}+k_{2}) =
\sigma^{(\infty)}(s,a_{1},a_{2}),\label{symmetry} \end{eqnarray} where
$k_{1}$ and $k_{2}$ are integers.
Also the effective potential of $A_i$ in abelian and non-abelian gauge
theories possesses the same periodicity and reaches its minima
(in our notation) at $a_i=n_i$ \cite{hosotani,weiss,hosotani1}. (see
also \cite{actor,smilga}). For further investigation we suppose that we
consider the sector with $0\leq a_i <1$ ($i=1,2$).
Moreover, one can check that \[
\sigma^{(\infty)}(s,a_{1},1-a_{2}) =   \sigma^{(\infty)}(s,1 -
a_{1},a_{2}) =    \sigma^{(\infty)}(s,a_{1},a_{2}), \] so that
$\sigma^{(\infty)}$ is symmetric with respect to the $a_{i}=1/2$. Thus
it is enough to consider the interval of values \begin{equation} 0
\leq a_{i} \leq 1/2, \; \; \; \; i=1,\  2.    \label{k17}
\end{equation} Let us mention that the special values $a_{1}=a_{2}=0$,
$a_{1}=a_{2}=1/2$ and $a_{1}=0$, $a_{2}=1/2$ represent respectively
the periodic (the case considered in Sects. 2 and 3), antiperiodic (or
twisted) and mixed (i.e. periodic in one toroidal direction and
antiperiodic in another) boundary conditions for the scalar field in
the absence of abelian gauge fields.

Now, let us return to the issue of the light mode. For the $a_{i}$ to
belong to the interval (\ref{k17}) the gauge components $A_{i}$ must
be of the order of $L^{-1}$, and thus there will be terms of the order
of $L^{-2}$ in eq. (\ref{k15}). However, as soon as $a_{i} < 1/2$,
$M_{(0,0)}$ remains to be the lowest mass in the spectrum. This is
also seen in Fig.~3, where we show the spectrum of the states with a
few lowest quantum numbers $N=(n_1,n_2)$ as a function of $a=a_1=a_2$.
We consider here the interval $0\leq a <1$ in order to make apparent
the symmetry of $\sigma ^{(\infty )}$ described in
eq.~(\ref{symmetry}). Restricting the values of $a_{i}$ to the
interval (\ref{k17}), we take the zero mode $\phi_{(0,0)}$ to be the
lightest one. This is the mode whose contribution is subtracted to
obtain $R$, eq. (\ref{eq:delta-def}). The sector of this mode appears
in the zero energy limit of the complete multidimensional theory.
Indeed, the difference between the zero mode mass and the next mass in
the spectrum is given by \[   M_{(1,0)}^2 - M_{(0,0)}^2 =
\frac{1-2a}{L^2} \rightarrow \infty \] as $L \rightarrow 0$ for
$a<1/2$. One also could think of $m_{0}^2$ being adjusted such that
$M_{(0,0)}^2 L^2 \ll 1$.

For numerical computations we take $a_{1}=a_{2}=a$. The behaviour of
$R$ as a function of $z=s/(4 M_{(1,0)}^2 (a=0)$ in the interval $0
\leq z \leq 1$ for various values of $a$ is plotted in Fig.~4. It is
described by the formula \begin{equation} R(z;a) = \frac{4}{9}
Z_2^{c^2}(2;a) z^2 +          \frac{8}{105} Z_2^{c^2}(3;a) z^3 + {\cal
O} (z^4), \label{ratio2} \end{equation} (cf. (\ref{ratio1})) valid for
$z < M^{2}_{(1,0)}(a)/M^{2}_{(1,0)}(a=0)$. Here \begin{equation}
Z_2^{c^2}(\nu;a)={\sum_{n_{1},n_{2}=-\infty}^{\infty}\!\!\!\!\!\!^{\prime}}
[(n_{1}-a)^2+(n_{2}-a)^2+c^2]^{-\nu}   \label{z-a} \end{equation} and
we supressed the dependence of $R$ on $w$ and instead indicated
explicitly its dependence on the parameter $a$. Again we see that the
behaviour of the deviation $R$ as a function of the parameter $s/(4
M_{(1,0)}(a)$ (which is not the same as $z$ !) is determined mainly by
the multiplicity $m_{1}(a)$ of the first heavy mass $M_{(1,0)}(a)$
which is now the function of $a$. From Fig. 3 one sees that
$m_{1}(0)=4$, $m_{1}(a)=2$ for $0 < a < 0.5$ and $m_{1}(0.5)=3$.

Let us consider the interval $0<z<1/2$. The curves in Fig. 4 show that
the function $R(z;a)$ for fixed $z$ grows with $a$ due to the increase
of the coefficient $(4/9) Z_2^{c^2}(2;a)$ of the first term in eq.
(\ref{ratio2}) with $a$. This can be understood as the result of the
two competing effects: the growth due to the approach of the first
heavy threshold and the decrease due to decrease of $m_{1}(a)$ when
$a$ departs from zero. The first effect wins in this competition. In
addition $m_{1}(a)$ increases when $a$ approaches the value $0.5$. As
the result, for example, $R(0.25;0.5)/R(0.25;0) \approx 2.6$ and
$R(0.5;0.5)/R(0.5;0) \approx 5.9$. This increase is seen clearer in
Fig. 5.

\section{Conclusions}

In continuation of the investigation carried out in
\cite{torus-scat,sphere-scat} in this article we considered the
$\lambda \phi^4$-theory on the space-time $M^4 \times T^2$. We studied
the scattering of two light particles in this model and calculated the
function $R$ which characterizes the deviation of the total cross
section of this process from the cross section of the same one but in
the $\lambda \phi^4$-model on $M^4$. The deviation is due to the
one-loop contributions of the heavy Kaluza-Klein modes, which appear
because of the multidimensional nature of the theory. For the centre
of mass energy $\sqrt{s}$ of the colliding particles below the
threshold of the first heavy particle the deviation grows with $s$ and
is already quite noticeable for $s > 0.25 \times$(energy of the first
threshold).

The behaviour of the function $R$ below the first threshold is given
with a good accuracy by the leading terms in the expansions
(\ref{ratio1}) or (\ref{ratio2}). Our results can be easily
generalized to an arbitrary two-dimensional compact manifold $K$ of
extra dimensions, then the formula for $R$ takes the form:
\begin{equation}    R \approx \frac{1}{36} \zeta (2 |K) (s L^{2})^{2}.
 \label{ratio3} \end{equation} Here $L$ is a scale such that the
eigenvalues of the Laplace operator are given by $\lambda_{N}/L^{2}$
and $\zeta(s | K)$ is the zeta-function of this operator: \[   \zeta(s
| K) = \sum_{\lambda_{N} \neq 0} \frac{m_{\lambda_{N}}}
{(\lambda_{N})^{s}}, \] where $m_{\lambda_{N}}$ is the multiplicity of
the eigenvalue $\lambda_{N}$. (Compare this with eqs. (\ref{k9}) and
(\ref{z-a}).) This formula was first obtained in ref.
\cite{sphere-scat} for the case of the sphere $K=S^{2}$. The
representation (\ref{ratio3}) tells that the behaviour of $R$ below
the first threshold is mainly determined by its position and the
multiplicity of the first heavy mass, which in their turn are
determined by the geometry of $K$. We demonstrated this in more detail
in the case of the non-equilateral torus and for the model with the
abelian gauge potential.

In the latter case we also have shown that for "low" energies of the
scattering particles, namely for $s L^{2} < 2 ( 1 + m_{0}^{2} L^{2})
\approx 2$, the function $R$ grows significantly (from $3$ to $6$
times)
when the angular variable $a = A_{i}L$, characterizing the gauge
potential, runs through the half-interval of periodicity, which ranges
from $0$ to $0.5$. Again the effect can be understood from the formula
(\ref{ratio3}): the interaction of the scalar field with the classical
constant (also slow varying on $M^{4}$) gauge potential produces the
change of the spectrum (masses and their multiplicities) of the
particles of the Kaluza-Klein tower, which in its turn leads to the
growth of $\zeta (2|K)$ with $a$.

We would like to mention that this effect of amplification of the
cross section might be relevant for the cosmology of the Early
Universe. Of course for $K=T^{2}$, though constant potential $A_{i}$
satisfies the equations of motion, such solution is not of much
physical interest, as is known. However, it seems that the effect of
amplification might take place in the case of more interesting gauge
models with the spaces of extra dimensions with non-zero curvature.
There the growth of the cross section due to the presence of
non-trivial gauge configurations, assuring spontaneous
compactification \cite{compact,KK-review2}, might give rise to
interesting consequences.

A few more remarks are in order here. There is a certain relation
between the effective potential of the constant gauge configuration
$A_{i}$ on $T^2$ (\cite{weiss,hosotani1}, see also \cite{actor,smilga})
and the
scattering of light scalar particles in this background, characterized
by the deviation $R$. The effective potential has its minima at $A_{i}
= n_{i}/L$. In the sector $0 \leq a \leq 0.5$, where we calculated
$R$, the minimum of the effective potential is reached
at $A_{1}=A_{2}=0$ for which the value of $R$ is minimal for a given
energy $\sqrt{s} < 2 M_{(1,0)}$ and the first heavy mass $M_{(1,0)}$
is maximal. Opposite to this, scattering of light scalar particles in
the background with $a=0.5$, corresponding to a maximum of the
effective potential, is characterized by the maximal value of the
deviation $R$ at low energies. It would be interesting to gain deeper
understanding of this relation. Other interesting physical effects in
abelian and non-abelian gauge theories with all or a part of the
space-time dimensions compactified to the torus (like confinement,
crossover and phase transitions, breaking of the gauge symmetry,
etc.) were studied in a number of papers \cite{decoupl},
\cite{gauge-torus}-\cite{polyakov}.

The model considered here is not a realistic one. Our aim was to
demonstrate the existence of the effect due to extra dimensions in the
behaviour of the total cross section of the particles of the low
energy sector of the theory, which is the $\lambda \phi^4$-model in
four dimensions in our case, and to study some characteristic features
of this effect. The function analogous to $R$ calculated in a
realistic extention of the Standard Model is to be compared with
experiment at future colliders. This could give an evidence of
existence of the Kaluza-Klein states or, by using plots like in Fig.
2, provide upper limits on the compactification scale $L$. We should
note that calculations of some processes for a certain class of
superstring models were carried out in \cite{antoniadis}.

\vspace{5mm}

\noindent{\large \bf Acknowledgments} We would like to thank Dom\'enec
Espriu, Andrei Smilga and Joan Soto for valuable discussions and
comments and the Department d'ECM of Barcelona University for the warm
hospitality. KK  acknowledges financial support from the Alexander von
Humboldt Foundation (Germany). YK acknowledges financial support from
CIRIT (Generalitat de Catalunya). 

\newpage

\section*{Figure captions}

\begin{description}   \item[Fig. 1] Plots of the ratio $R$, eq.
         (\ref{eq:delta-def}), as a function of $z= s/(4
M_{(1,0)}^2)$, where $M_{(1,0)}^2 = m_{0}^2 + 1/L_{1}^2$, for various
values of $w=(L_{1}/L_{2})^2$ in the interval $0 \leq z \leq 1$.
\item[Fig. 2] Plots of $R$ as a function of $x= 1/(L m_{0})$ for
          the fixed energy given by $r = s/m_{0}^{2}=80$ and $r=400$.
The case of the equilateral torus.   \item[Fig. 3] The spectrum
(\ref{k14}) of the masses of the                 states with lowest
quantum numbers $N=(n_{1},n_{2})$ as functions of $a=a_{1}=a_{2}$,
where $a_{i}=A_{i}L$, $i=1,2$. $H(N) \equiv M_{N}^{2}(a) - m_{0}^2 =
(n_{1}-a)^{2}+(n_{2}-a)^{2}$.   \item[Fig. 4] Plots of the ratio $R$
for                 the $\lambda \phi^4$-model with abelian gauge
field as a function of $z=s/[4M_{(1,0)}(a=0)]$ for various values of
$a=a_{1}=a_{2}$. Here $L_{1}=L_{2}=L$.    \item[Fig. 5] Plots of $R$
as a function of $a=a_{1}=a_{2}$                 for fixed values of
$z= s/(4 M_{(1,0)}^{2}(a=0))$. $L=L_{1}=L_{2}$. \end{description}


\bigskip

\newpage

\begin{thebibliography}{99}

\bibitem[*]{hugo} Alexander von Humboldt foundation fellow. E-mail
address: klaus@zeta.ecm.ub.es.

\bibitem[\dag]{yk} On leave of absence from Nuclear Physics Institute,
Moscow State University, 117234 Moscow, Russia. E-mail address:
kubyshin@ecm.ub.es.

\bibitem[1]{torus-scat} A.P. Demichev, Yu.A. Kubyshin and J.I. P\'erez
Cadenas, {\em Phys. Lett.} {\bf B323} (1994) 139.

\bibitem[2]{sphere-scat} E. Elizalde and Yu. Kubyshin. {\em Possible
evidences of Kaluza-Klein particles in a scalar model with spherical
compactification} Preprint UAB-ECM-PF 94/4, University of Barcelona
(1994), to appear in J. Phys. A.

\bibitem[3]{top-mass} G.W. Gibbons, {\em J. Phys.} {\bf A11} (1978)
1341. \\ E.J. Copeland and D.J. Toms, {\em Nucl. Phys.} {\bf B255}
(1985) 201. \\ G.R. Shore, {\em Ann. Phys.} {\bf 128} (1980) 376. \\
G. Kennedy, {\em Phys. Rev.} {\bf D23} (1981) 2884. \\ D.J. Toms, {\em
Phys. Rev.} {\bf D25} (1982) 2536. \\ Yu.P. Goncharov, {\em Phys.
Lett.} {\bf 119B} (1982) 403. \\ J.S. Dowker, {\em Class. Quant.
Grav.} {\bf 1} (1984) 359. \\ A. Actor, {\em Class. Quant. Grav.} {\bf
7} (1990) 1463. \\ G. Cognola, K. Kirsten and S. Zerbini, {\em Phys.
Rev.} {\bf D48} (1993) 790.\\ E. Elizalde and A. Romeo, {\em Phys.
Lett.} {\bf B244} (1990) 387.\\ G. Denardo and E. Spallucci, {\em
Nuovo Cimento} {\bf A64} (1981) 27.\\ G. Denardo and E. Spallucci,
{\em Nucl. Phys.} {\bf B169} (1980) 514.\\ K. Kirsten, {\em Class.
Quantum Grav.} {\bf 11} (1994) 57.\\ A.A. Bytsenko, G. Cognola, L.
Vanzo and S. Zerbini, {\em Quantum Fields and Extended Objects in
Space-Times with Constant Curvature Spatial Section}, Preprint Trento,
U.T.F. 325.

\bibitem[4]{eekk} E. Elizalde and K. Kirsten, {\em J. Math. Phys.}
{\bf 35} (1994) 1260.

\bibitem[5]{kk} K. Kirsten, {\em J. Phys.} {\bf A26} (1993) 2421.

\bibitem[6]{ambjorn} J. Ambjorn and S. Wolfram, {\em Ann. Phys.} {\bf
147} (1983) 1.

\bibitem[7]{Casimir} H.B.G. Casimir, {\em Proc. Kon. Ned. Akad. Wet.}
{\bf 51} (1948) 793.\\ S.K. Blau, M. Visser and A. Wipf, {\em Nucl.
Phys.} {\bf B310} (1988) 163.\\ E. Elizalde, {\em Nuovo Cimento} {\bf
B104} (1989) 685.\\ G. Cognola, L. Vanzo and S. Zerbini, {\em J. Math.
Phys.} {\bf 33} (1992) 222.\\ F. Caruso, N.P. Neto, B.F. Svaiter and
N.F. Svaiter, {\em Phys. Rev.} {\bf D43} (1991) 1300.\\ B.F. Svaiter
and N.F. Svaiter, {\em J. Math. Phys.} {\bf 32} (1991) 175.\\ L.H.
Ford, {\em Phys. Rev.} {\bf D21} (1980) 933.\\ D.J. Toms, {\em Phys.
Rev.} {\bf D21} (1980) 2805.\\ D.J. Toms, {\em Phys. Rev.} {\bf D21}
(1980) 928.\\ K. Kirsten, {\em J. Phys. A} {\bf A24} (1991) 3281.\\ K.
Kirsten, {\em Class. Quantum Grav.} {\bf 8} (1991) 2239.\\ J.S. Dowker
and J.P. Schofield, {\em Nucl. Phys.} {\bf B327} (1989) 267.\\ J.S.
Dowker and R. Banach, {\em J. Phys.} {\bf A11} (1978) 2255. \\ P.
Candelas and S. Weinberg, {\em Nucl. Phys.} {\bf B237} (1984) 397. \\
B.P. Dolan and C. Nash, {\em Commun. Math. Phys.} {\bf 148} (1992)
139.

\bibitem[8]{temperature} C.W. Bernard, {\em Phys. Rev.} {\bf D9}
(1974) 3312.\\ L. Dolan and R. Jackiw, {\em Phys. Rev.} {\bf D9}
(1974) 3320.\\ N.P. Landsman and Ch.G. von Weert, {\em Phys. Rep.}
{\bf 145} (1987) 141.\\ D.J. Gross, R.D. Pisarski and S. Rudaz, {\em
Rev. Mod. Phys.} {\bf 53} (1981) 43.\\ J.I. Kapusta, {\em Finite
temperature field theory} (Cambridge University, Cambridge, 1989).

\bibitem[9]{DIKT} A.P. Demichev, M.Z. Iofa, Yu.A. Kubyshin and V.E.
Tarasov, {\em Sov. Yad. Fis.} {\bf 56} (1993) 222.

\bibitem[10]{antoniadis} I. Antoniadis and K. Benakli, {\em Phys.
Lett.} {\bf B326} (1994) 69. \\ I. Antoniadis, K. Benakli and M.
Quir\'os, {\em Phys. Lett.} {\bf B331} (1994) 313.

\bibitem[11]{decoupl} Yu. Kubyshin, D. O'Connor and C.R. Stephens,
{\em Class. Quant. Grav.} {\bf 10} (1993) 2519; Yu. Kubyshin, D.
O'Connor and C.R. Stephens, in the Proc. of the ``Quarks-92" Seminar.

\bibitem[12]{compact} E. Cremmer and J. Scherk, {\em Nucl. Phys.} {\bf
B118}(1977) 61. \\ J.F. Luciani, {\em Nucl. Phys.} {\bf B135} (1978)
11.

\bibitem[13]{KK-review2} Yu.A. Kubyshin, J.M. Mour\~{a}o, G. Rudolph
and I.P. Volobujev, {\em Dimensional Reduction of Gauge Theories,
Spontaneous Compactification and Model Building}, Lecture Notes in
Physics, Vol. 349 (Springer-Verlag, Berlin, 1989). \\ D. Kapetanakis
and G. Zoupanos, {\em Phys. Rep.} {\bf C219} (1992) 1.

\bibitem[14]{kolb} E.W. Kolb and R. Slansky, {\em Phys. Lett.} {\bf
135B} (1984) 378.

\bibitem[15]{kapl-88} V.S. Kaplunovsky, {\em Nucl. Phys.} {\bf B307}
(1988) 145. \\ L.J. Dixon, V.S. Kaplunovsky and J. Louis, {\em Nucl.
Phys.} {\bf B355} (1991) 649.\\ I. Antoniadis, {\em Phys. Lett.} {\bf
B246} (1990) 377.

\bibitem[16]{hawk} D. Ray and I. Singer, {\em Adv. Math.} {\bf 7}
(1971) 145. \\ R. Critchley and J.S. Dowker, {\em Phys. Rev.} {\bf
D13} (1976) 3224. \\ S. Hawking, {\em Commun. Math. Phys.} {\bf 55}
(1977) 133.

\bibitem[17]{hille} E. Hille, {\em Analytic Function Theory}, Vol. II,
Ginn, Boston, (1962).

\bibitem[18]{grad} I.S. Gradshteyn and I.M. Ryzhik, {\em Tables of
Integrals, Series and Products}, Academic Press, New York, (1965).

\bibitem[19]{gauge-torus} G. 't Hooft, {\em Nucl. Phys.} {\bf B153}
(1979) 141; {\em Acta Phys. Austriaca Suppl.} {\bf XXII} (1980).  \\
S. Cecotti and L. Girardelo, {\em Nucl. Phys.} {\bf B208} (1982) 265.

\bibitem[20]{actor} A. Actor, {\em J. Math. Phys.} {\bf 25} (1984)
2736; {\em Ann. Phys.} {\bf 159} (1985) 445.

\bibitem[21]{smilga} A.V. Smilga, {\em Ann. Phys.} {\bf 234} (1994) 1.

\bibitem[22]{angular} E. Gava, R. Jengo and C. Omero, {\em Nucl.
Phys.} {\bf B170} (1980) 445. \\ L. G. Yaffe and B. Svetitsky, {\em
Phys. Rev.} {\bf D26} (1982) 963. \\ D. J. Gross, R.D. Pisarski and L.
G. Yaffe, {\em Rev. Mod. Phys.} {\bf 53} (1981) 43.

\bibitem[23]{hosotani}
Y. Hosotani, {\em Phys. Lett. B} {\bf 126} (1983) 309.\\
B. Svetitsky and L. G. Yaffe, {\em Nucl. Phys. B} {\bf 210} (1982)
423.\\
D.J. Toms, {\em Phys. Lett. B} {\bf 126} (1983) 445.\\
N.S. Manton, {\em Ann. Phys.} {\bf 159} (1985) 220.

\bibitem[24]{weiss} N. Weiss, {\em Phys. Rev.} {\bf D24} (1981) 475;
{\bf D25} (1982) 2667.

\bibitem[25]{hosotani1}
C.-L. Ho and Y. Hosotani, {\em Nucl. Phys. B} {\bf 345} (1990) 445.

\bibitem[26]{polyakov} A.M. Polyakov, {\em Phys. Lett.} {\bf B72}
(1978) 477. \\ L. Susskind, {\em Phys. Rev.} {\bf D20} (1979) 2610. \\
L.D. McLerran and B. Svetitsky, {\em Phys. Rev.} {\bf D24} (1981) 450.

\end{thebibliography}
\end{document}